\documentstyle[aps,12pt,manuscript]{revtex}
\begin{document}
\preprint{INJE-TP-98-5 \& hep-th/9805050}

\title{Dilaton Test of Connection between AdS$_3 \times $S$^3$ and 
5D black hole}

\author{ H.W. Lee, N.J. Kim and Y. S. Myung}
\address{Department of Physics, Inje University, Kimhae 621-749, Korea} 

\maketitle

\begin{abstract}
A 5D black hole(M$_5$) is investigated in the type IIB superstring theory 
compactified on S$^1 \times $T$^4$. 
This corresponds to AdS$_3 \times $S$^3 \times $T$^4$ in the near
horizon with asymptotically flat space.
Here the harmonic gauge is introduced to decouple the mixing 
between the dilaton and others.  
On the other hand we obtain the BTZ balck 
hole(AdS$_3\times$S$^3\times$T$^4$) as the non-dilatonic solution.
We calculate the greybody factor  
of the dilaton as a test scalar both for a 5D black hole(
M$_5 \times $S$^1 \times $T$^4$) 
and the BTZ black hole(AdS$_3 \times $S$^3 \times $T$^4$). 
The result of the BTZ black hole agrees with the greybody factor of the dilaton 
in the dilute gas approximation of a 5D black hole.
\end{abstract}

\newpage
\section{Introduction}
\label{introduction}
Recently anti-de Sitter spacetime(AdS) has attracted much interest. It
appears that the conjecture relating
the string(supergravity) theory on AdS to conformal field theory(CFT) 
on its boundary
may resolve many problems in black hole 
physics\cite{Mal98ATMP231,Gub98PLB105,Wit98ATMP253,Hor98PRL4116}.
The 5D black hole correspond to U-dual to
BTZ$\times $S$^3$\cite{Hyu97,Sfe98NPB179}. 
Sfetsos and Skenderis calculated the entropies
of non-extremal 5D black holes by applying Carlip's
approach to the BTZ black hole. The BTZ black hole(locally AdS$_3$)
has no curvature singularity\cite{Ban92}
and is considered as a prototype for the general AdS/CFT correspondence.
This is actually an exact solution of string theory \cite{Hor93}
and there is an exact CFT with it on the
boundary. Carlip has shown that the physical boundary
degrees of freedom account for the Bekenstein-Hawking
entropy of the BTZ black hole\cite{Car95}.
The other calculation was found in \cite{Str97,Beh98PLB310}.

More recently, two of authors (Lee and Myung) calculated the 
greybody factor(absorption cross section) for a minimally coupled 
scalar both in M$_5 \times $S$^1 \times $T$^4$ and 
AdS$_3 \times $S$^3 \times $T$^4$\cite{Lee98}.  
These geometries are two solutions in type IIB theory: the first is the 
dilatonic solution(a 5D black hole), while the second is the 
non-dilatonic solution(the BTZ black hole).
Although the near horizon geometries of two solutions are the same as 
AdS$_3$, the asymptotic structures are quite different.
It turned out that two results are the same.  This means that the near horizon 
AdS$_3$-geometry of a 5D black hole contains 
the essential information of a 5D black hole.  
However, this is the result for a free scalar, 
which decouples to any other fields.  
In order to confirm 
this result, we need an another test field 
which is a fixed scalar(a 10D dilaton).  

In this paper we wish to calculate the greybody factor, using the dilaton. 
It is worth noting that an absorption cross section can be 
extracted from a solution of the differential equation.  
The perturbation analysis around M$_5 \times $S$^1 \times $T$^4$ and
AdS$_3 \times $S$^3 \times $T$^4$ reveals a complicated 
mixing between the dilaton and 
other fields.  Here introducing the harmonic gauge, we disentangle 
the mixing completely and thus obtain decoupled dilaton equations.  
Sec.\ref{sugra} deals with 
the supergravity calculation in a 5D black hole 
background. 
In Sec.\ref{ads} we find the BTZ solution and 
perform the same calculation to obtain the greybody factor of the dilaton 
in AdS$_3 \times $S$^3 \times $T$^4$.
Here we do not require any (Dirichlet or Neumann) condition 
at the spatial infinity, but the non-normalizable mode is used 
as a boundary condition to 
calculate the greybody factor.
Finally we discuss our results in Sec.\ref{discussion}. 
In Appendix \ref{RMN}, the explicit form of ${\bar R}_{MN}$ in the 
Einstein frame is presented.  The form of the graviton $h_{MN}$ is 
given in Appendix \ref{hMN}.

\section{Supergravity Calculation : Scattering in 5D black hole}
\label{sugra}
\subsection{Background Solution}
The low-energy action of type IIB theory in the Einstein frame 
is given by\cite{Hor96}
\begin{equation}
S^{10}_{\rm E} = {1 \over 16 \pi G^{10}_N} 
\int d^{10} x \sqrt{-g} \left [ R - {1 \over 2} (\nabla \Phi)^2 
- { 1\over 12} e^{\Phi} H^2 \right ].
\label{einstein-action}
\end{equation}
Here $H$ denotes the RR-three form and $\Phi$ is the 10D dilaton.
The equations of motion for the action (\ref{einstein-action}) lead to 
\begin{eqnarray}
&&R_{MN} =  {1 \over 2} \nabla_M \Phi \nabla_N \Phi 
           + {1 \over 4} e^{\Phi} H_{MPQ} H_N^{~~PQ}
           - {1 \over 48} e^{\Phi} H^2 g_{MN},
\label{eq-Rmn}\\
&&\nabla^2 \Phi -{1 \over 12} e^{\Phi} H^2 =0,
\label{eq-Phi}\\
&&\nabla_M(e^\Phi H^{MNP} ) =0.
\label{eq-H}
\end{eqnarray}
In addition, we have one Bianchi identity as the remaining 
Maxwell's equations
\begin{equation}
\partial_{[M} H_{NPQ]} = 0.
\label{bianchi-identity}
\end{equation}
The contraction form of Eq.(\ref{eq-Rmn}) is given by
\begin{equation}
R-{1 \over 2} (\nabla \Phi)^2 - {1 \over 2} \nabla^2 \Phi =0.
\label{eq-contract}
\end{equation}
The solution to Eqs.(\ref{eq-Rmn})-(\ref{eq-H}) are given by a 5D black hole 
and additional Kaluza-Klein moduli ($\nu, \nu_5$)
\begin{eqnarray}
e^{-2 \bar \Phi} &=& {1 \over g^2} { f_5 \over f_1}, ~~
\bar H = { {2 r_5^2} \over g} \epsilon_3 + 
2 r_1^2 g e^{-2 \bar \Phi} *_6 \epsilon_3, 
\label{bck-sol} \\
\left ( ds_{10}^2 \right )_E &=& 
e^{2 \bar \nu} dx_i^2 +
  e^{2 \bar \nu_5} ( dx_5 + A_{\tilde P} dx^{\tilde P})^2 +
  e^{-2(4 \bar \nu + \bar \nu_5)/3} ds_{5D}^2
\label{kaluza-metric}
\end{eqnarray}
with $\tilde P = 0,1,2,3,4$.
This corresponds to the dilatonic solution.
Here $e^{2 \bar \nu} = f_1^{1/4} f_5^{-1/4},~
e^{2 \bar \nu_5} = f_1^{-3/4} f_5^{-1/4} f_n$, 
and the Kaluza-Klein gauge potential 
$A_0 = -r_0^2 \sinh 2 \sigma / 2(r^2+r_n^2)$. 
A 5D black hole space-time(M$_5$) is given by 
\begin{equation}
ds_{5D}^2 = - d f^{-2/3} dt^2 + f^{1/3} [ d^{-1} dr^2 + r^2 d \Omega_3^2]
\label{5d-metric}
\end{equation}
with $f = f_1 f_5 f_n$, $f_i = 1 + r_i^2 / r^2$ and $d = 1 - r_0^2/r^2$.
For simplicity, we consider only the dilute gas approximation of 
$r_1 = r_5 = R \gg r_0,r_n$ and $A_0 \simeq 0$.  In this case one finds 
M$_5 \times $S$^1 \times $T$^4$ and its near horizon geometry with 
$\bar \Phi={\rm ~constant},~~ \bar R \simeq 0,~~ \bar H^2 =0$ 
takes AdS$_3\times$S$^3\times$T$^4$ as 
\begin{equation}
\left ( ds_{10}^2\right )_E \simeq 
ds_{BTZ}^2 + R^2 d\Omega_3^2 + {r_1^2 \over R^2} dx_i^2,
\label{BTZ-metric}
\end{equation}
where the BTZ black hole space-time is given by\cite{Ban92}
\begin{equation}
ds_{BTZ}^2  = - {{(\rho^2 - \rho_+^2)(\rho^2 -\rho_-^2)} \over
   {\rho^2 R^2 }} dt^2 + \rho^2 ( d \varphi - {J \over 2 \rho^2} dt )^2 +
  { \rho^2 R^2 \over {(\rho^2 - \rho_+^2)(\rho^2 -\rho_-^2)}} d \rho^2.
\label{BTZ-metric3}
\end{equation}
The explicit form of $\bar R_{MN}$ is given in Appendix \ref{RMN}.

\subsection{Perturbation}
Now we introduce the small perturbations around the dilute gas 
background as
\begin{eqnarray}
g_{MN} &=& \bar g_{MN} + h_{MN},
\label{ptr-metric} \\
H_{MNP} &=& \bar H_{MNP} + {\cal H}_{MNP},
\label{ptr-H} \\
\Phi &=& \bar \Phi + \delta \phi,
\label{ptr-Phi}
\end{eqnarray}
where $h_{MN}$ is given in Appendix \ref{hMN}.  This is 
enough for the s-wave calculation.  
Then the linearized equations for Eqs.(\ref{eq-contract}), 
(\ref{eq-Phi}), (\ref{eq-H}) take the form
\begin{eqnarray}
&&\bar g^{MN} \delta R_{MN}(h) 
  - h^{MN} \bar R_{MN} 
  - {1 \over 2} \bar \nabla^2 \delta \phi = 0,
\label{eq-Rmn-ptr}\\
&&\bar \nabla^2 \delta \phi 
-{1 \over 12} \left \{ 2 \bar H_{MNP} {\cal H}^{MNP} 
  - 3 \bar H_{MNP} \bar H^{QNP} h^M_{~Q} 
  \right \} =0,
\label{eq-Phi-ptr}\\
&&\bar \nabla_M{\cal H}^{MNP} 
 - (\bar \nabla_M h^N_{~Q}) \bar H^{MQP} 
 + (\bar \nabla_M h^P_{~Q}) \bar H^{MQN} 
\nonumber \\
&&~~~~~~~~~
 - (\bar \nabla_M \hat h^M_{~Q}) \bar H^{QNP} 
 - h^M_{~Q} (\bar \nabla_M  \bar H^{QNP}) 
 + (\partial_M \delta \phi) \bar H^{MNP} 
=0,
\label{eq-H-ptr}
\end{eqnarray}
where the Lichnerowitz operator $\delta R_{MN}(h)$ and $\hat h_{MN}$ 
are given by\cite{Gre93} 
\begin{eqnarray}
\delta R_{MN}(h) &=& - {1 \over 2} \bar \nabla^2 h_{MN} 
   + \bar R_{Q(M}h^Q_{~N)} - \bar R_{PMQN}h^{PQ} 
   - \bar \nabla_{(M} \bar \nabla_{|P|}\hat h^P_{~N)},
\label{del-Rmn} \\
\hat h_{MN} &=& h_{MN} - { h \over 2} \bar g_{MN},~~h = h^Q_{~Q}.
\label{hat-h}
\end{eqnarray}
Here for simplicity, one chooses the harmonic gauge for the graviton,
\begin{equation}
\bar \nabla_M \hat h^{MP} = 0.
\label{harmonic}
\end{equation}
For the s-wave calculation, we need not the full linearized equation
($\delta R_{MN} \cdots $)  of Eq.(\ref{eq-Rmn}) 
but its scalar equation ($\delta R \cdots $) of Eq.(\ref{eq-contract}).
Eqs. (\ref{eq-Rmn-ptr}) and (\ref{eq-Phi-ptr}) lead to
\begin{eqnarray}
&&\bar \nabla^2 (h+\delta \phi) 
- {4 R^4 d \over r^6 f_5^3} ( h^r_{~r} +  h^t_{~t} 
+ h^{x_5}_{~x_5} -  h^{\theta_i}_{~\theta_i})=0,
\label{eq-h-ptr} \\
&&\bar \nabla^2 \delta \phi 
+ {4 R^4 \over r^6 f_5^3} \left [ 2({\cal H} -{\cal H}_\theta ) -
(h^r_{~r} +  h^t_{~t} 
+ h^{x_5}_{~x_5} -  h^{\theta_i}_{~\theta_i}) \right ]=0.
\label{eq-phi-ptr} 
\end{eqnarray}
Here ${\cal H}$ and ${\cal H}_\theta$ are defined as 
\begin{equation}
{\cal H}_{\epsilon_3} = {\cal H}_{\theta} 
             \left ({ 2 r_5^2 \over g} \epsilon_3\right ), ~~
{\cal H}_{*_6\epsilon_3} = {\cal H} 
\left (2 r_1^2  g e^{-2 \bar\Phi}*_6 \epsilon_3\right ).
\label{def-curlH}
\end{equation}
Now we attempt to disentangle the last terms in Eq.(\ref{eq-h-ptr}) and 
(\ref{eq-phi-ptr}) by using both the harmonic gauge and Kalb-Ramond equation. 
When $N=t, P=x_5$, solving both Eqs. (\ref{eq-H-ptr}) and (\ref{harmonic}) 
leads to
\begin{eqnarray}
\left ({\cal H} + \delta \phi  -h^t_{~t} -  h^{x_5}_{~x_5} -  h^r_{~r}
  + { h \over 2} \right )' =0,
\label{constraint}
\end{eqnarray} 
where the prime($'$) means the differentiation with respect to $r$.  
Here one finds an important relation
\begin{equation}
2 {\cal H} + 2 \delta \phi  -h^t_{~t} -  h^{x_5}_{~x_5} -  h^r_{~r}
+h^{\theta_i}_{~\theta_i} + h^i_{~i} =0.
\label{relation1}
\end{equation}
The remaining choices of $N,P$ lead to the same relation (\ref{relation1}).
For $N=\theta_2, P=\theta_3$, one obtains the relation
\begin{equation}
2 {\cal H}_{\theta} + 2 \delta \phi  +h^t_{~t} +  h^{x_5}_{~x_5} +  h^r_{~r}
-h^{\theta_i}_{~\theta_i} + h^i_{~i} =0.
\label{relation2}
\end{equation}
On the other hand, if one uses the dilaton gauge
($h^{RQ} \Gamma^P_{RQ} = \bar \nabla_R \hat h^{RP}$) 
in Refs.\cite{Lee9801,Lee98PRD084022}, instead of 
the harmonic gauge, we obtain the same relations (\ref{relation1}) and 
(\ref{relation2}).
Further taking into account the Bianchi identity of 
Eq.(\ref{bianchi-identity}) leads to
\begin{eqnarray*}
{\cal H} &=& {\cal H}(t,x_5,r) \Rightarrow ~~{\rm a~ dynamical~ one},
\\
{\cal H}_{\theta} &=& {\cal H}_{\theta}(\theta_1,\theta_2,\theta_3) 
\Rightarrow ~~{\rm a~ non~dynamical~ one}.
\end{eqnarray*}
And thus we can set ${\cal H}_\theta$ to be zero.  
Using Eqs.(\ref{relation1})-(\ref{relation2}), Eqs.
(\ref{eq-h-ptr}) and (\ref{eq-phi-ptr}) reduce to
\begin{eqnarray}
&&\bar \nabla^2 (h+\delta \phi) 
+ {4 R^4 d \over r^6 f_5^3} ( 2 \delta \phi +  h^i_{~i} ) =0,
\label{eq-h-ptr1} \\
&&\bar \nabla^2 \delta \phi 
- {2 R^4 \over r^6 f_5^3} ( 2\delta \phi  +  h^i_{~i}  ) =0.
\label{eq-phi-ptr1} 
\end{eqnarray}
Hereafter we consider only the propagation of the dilaton in 
Eq.(\ref{eq-phi-ptr1}). 
Assuming $h^i_{~i} = b \delta \phi$, equation 
(\ref{eq-phi-ptr1}) takes the form 
\begin{equation}
\bar \nabla^2 \delta \phi 
- {2 R^4 \over r^6 f_5^3} ( 2 +b) \delta \phi  =0.
\label{eq-phi-ptr2} 
\end{equation}
Now we are in a position to determine the parameter $b$.  
From (\ref{bck-sol}) and (\ref{kaluza-metric}) one gets 
the relation
\begin{equation}
e^{2 \bar \Phi} = g^2 { f_1 \over f_5} = g^2 e^{8 \bar \nu},
\label{phi-nu}
\end{equation}
which implies that the linearized relation for 
$\delta \nu$($\nu = \bar \nu + \delta \nu$) is given by
\begin{equation}
\delta \phi = 4 \delta \nu.
\label{phi-nu-ptr}
\end{equation}
Considering the perturbation $g_{ij} = \bar g_{ij} + h_{ij}$, then 
one obtains
\begin{equation}
h^i_{~i} = 8 \delta \nu = 2 \delta \phi,
\label{hii}
\end{equation}
which means that $b=2$.

As a result, one obtains the decoupled dilaton equation
\begin{equation}
\bar \nabla^2 \delta \phi 
- {8 R^4 \over r^6 f_5^3}  \delta \phi  =0.
\label{eq-dilaton} 
\end{equation}
In deriving the decoupled equation, we may choose either the dilaton 
gauge or the Krasnitz-Klebanov setting for $h_{\mu\nu}$\cite{Lee9708}.  
However, these calculations lead to the same equation (\ref{eq-dilaton}).  
This means that the dilaton is a gauge invariant quantity.  And thus 
it is regarded as a good test field.  This is a reason why we choose 
$h_{MN}$ to be block diagonal as in Appendix B.  

\subsection{Greybody Factor}
Taking into account the symmetries in (\ref{kaluza-metric}), 
$\delta \phi$ can be seperated as
\begin{equation}
\delta \phi = e^{-i \omega t} e^{i K_5 x_5} e^{i K_i x^i} Y_l(\theta_1, 
\theta_2, \theta_3) \delta \tilde \phi(r).
\label{Phi}
\end{equation}
First let us consider the wave equation in the ten-dimensional background 
(\ref{kaluza-metric})
\begin{equation}
\left [ {d^2 \over dr^2} + \left ( { d' \over d} + {3 \over r} \right ) 
 { d \over dr} \right ] \delta \tilde \phi + 
\left [ {\omega^2 f \over d^2}  
-{f_5^2 \over f_n d} K_5^2 
-{f_5 \over d} K_i^2 
-{l(l+2) \over r^2 d} 
- {8R^4 \over r^6f_5^2d} \right ] \delta \tilde \phi= 0.
\label{wave-eq}
\end{equation}
For s-wave calculation, we take $K_5=K_i=0,~l=0$. 
Then (\ref{wave-eq}) leads to the 
effective 5D wave equation
\begin{equation}
\left [ {d^2 \over dr^2} + \left ( { d' \over d} + {3 \over r} \right ) 
 { d \over dr} \right ] \delta \tilde \phi + 
{\omega^2 f \over d^2} \delta \tilde \phi 
- {8R^4 \over r^2(r^2+R^2)^2d} \delta \tilde \phi= 0.
\label{r-eq}
\end{equation}
This is just the linearized equation of the fixed scalar $\nu$ 
in the effective 5D theory\cite{Lee9708,Cal97}. 
The s-wave absorption cross section for M$_5 \times $S$^1$ 
is given by\cite{Cal97}
\begin{equation}
\left ( \sigma^{\delta \phi}_{M_5 \times S^1} \right )_{l=0,K_5=0} 
= \left ( {\pi^3 R^8 \over 64} \right )
2 \pi R ( \omega^2 + 16 \pi^2 T_R^2)
(\omega^2 + 16 \pi^2 T_L^2) \omega 
{ {e^{\omega/T_H} -1 } \over {[e^{\omega/2 T_R}-1][e^{\omega/2 T_L}-1]}}.
\label{abs-M5}
\end{equation}
In the low energy limit of $\omega\to 0(\omega<T_L, T_R, T_H)$, this leads to
\begin{equation}
\left (\sigma^{\delta \phi}_{M_5 \times S^1} \right )_{l=0,K_5=0}^{\omega\to 0} 
= {{\cal A}_H^{6D} \over 4} \left ( {r_0 \over R} \right )^4,
\label{abs-M5-limit}
\end{equation}
where ${\cal A}_H^{6D}(= {\cal A}_H^{5D} \times 2 \pi R = 4 \pi^3 r_n R^3)$ 
is the area of horizon for M$_5 \times $S$^1$.

\section{AdS-Calculation : Scattering in AdS$_3 \times $S$^3 \times $T$^4$}
\label{ads}
\subsection{Background Solution}
So far we calculate the cross section for a 5D black hole with
asymptotically flat space.
This leads to AdS$_3 \times $S$^3 \times $T$^4$ only in the near horizon.
However, the AdS$_3 \times $S$^3 \times $T$^4$ exists as 
the other solution to the type IIB theory
in the string frame. Hence we can extend the
near horizon AdS$_3$-geometry to the AdS-spacetime
as a whole. This new solution is useful for comparison.

We start with the action in the string frame
\begin{equation}
S^{10}_{\rm string} = { 1\over 16 \pi G^{10}_N}
\int d^{10}x \sqrt{-g} \left [ e^{-2 \Phi} \left ( R + 4 (\nabla \Phi)^2
\right ) - {1 \over 12} H^2 \right ]. 
\label{action-string}
\end{equation}
The equations of motion are given by
\begin{eqnarray}
&&R_{MN} =
           -{1 \over 4} \nabla^2 \Phi g_{MN}
           + {1 \over 2} (\nabla \Phi)^2 g_{MN}   
           -  2 \nabla_M \nabla_N \Phi 
           + {1 \over 4} e^{2\Phi} H_{MPQ} H_N^{~~PQ}
           - {1 \over 48} e^{2 \Phi} H^2 g_{MN}, \hspace*{-1.3pt}
\label{eq-Rmn-b}\\
&&R -  4 (\nabla \Phi)^2 + 4 \nabla^2 \Phi =0,
\label{eq-Phi-b}\\
&&\nabla_M(H^{MNP} ) =0.
\label{eq-H-b}
\end{eqnarray}
From Eqs.(\ref{eq-Rmn-b}) and (\ref{eq-Phi-b}), one finds the new 
dilaton equation
\begin{equation}
\nabla^2 \Phi - 2 (\nabla \Phi)^2 - {1 \over 12} e ^{2 \Phi} H^2 = 0.
\label{eq-dilaton-b}
\end{equation}
Solving the above equations, the BTZ black 
hole(AdS$_3 \times $S$^3 \times $T$^4$) is obtained as
\begin{eqnarray}
&&\bar \Phi = {\rm constant}, ~~\bar R = 0,~~\bar H^2 =0,
~~\bar R_{MN} = {1 \over 4} \bar H_{MPQ} \bar H_N^{~PQ},
\nonumber \\
&&\bar R_{\mu\nu}^{BTZ} = - {2 \over R^2} \bar g^{BTZ}_{\mu\nu}, ~~
\bar R^{S^3}_{\mu\nu} = {2 \over R^2} \bar g^{S^3}_{\mu\nu},
~~\bar H^{BTZ} = { 2 \over {gR}} *_6 \epsilon_3, 
~~\bar H^{S^3} = {2 \over {g R}} \epsilon_3.
\label{bck-b}
\end{eqnarray}
The dilaton($\Phi$) plays no role in making the black hole solution.
In this sense we call this as non-dilatonic solution.

\subsection{Perturbations}
The linearized equations for (\ref{eq-Phi-b}) and (\ref{eq-dilaton-b}) 
around the AdS$_3 \times $S$^3 \times $T$^4$ background take 
the form
\begin{eqnarray}
&&\bar \nabla^2 (8\delta \phi-h) 
+ {4 \over R^2 } ( h^t_{~t} + h^{\varphi}_{~\varphi} + h^{\rho}_{~\rho}  
 -  h^{\theta_i}_{~\theta_i})=0,
\label{eq-h-ptr-b} \\
&&\bar \nabla^2 \delta \phi 
+ {2  \over R^2 } \left [ 2({\cal H} -{\cal H}_\theta ) -
( h^t_{~t} + h^{\varphi}_{~\varphi} + h^{\rho}_{~\rho}
 -  h^{\theta_i}_{~\theta_i}) \right ]=0.
\label{eq-phi-ptr-b} 
\end{eqnarray}
Here $h_{MN}$ takes the similar form as in Appendix \ref{hMN}.  
In order to decouple the last terms in the above, we need the harmonic gauge 
 in Eq.(\ref{harmonic}) and Kalb-Ramond equation as \cite{Lee98PRD084022}
\begin{equation}
\bar \nabla_M {\cal H}^{MPQ} 
-\bar \nabla_M h_N^{~P} \bar H^{MNQ} 
+ \bar \nabla_M h_N^{~Q} \bar H^{MNP}
-(\bar \nabla_M \hat h^M_{~N}) \bar H^{NPQ}
-h^M_{~N} (\bar \nabla_M \bar H^{NPQ})
=0.
\label{kalb-b}
\end{equation}
From these one finds two relations for AdS$_3$ and S$^3$ 
\begin{eqnarray}
2 {\cal H} - ( h^t_{~t} + h^{\varphi}_{~\varphi} + h^{\rho}_{~\rho}) 
      +  h^{\theta_i}_{~\theta_i} + h^i_{~i} =0,
\label{relation1-b} \\
2 {\cal H}_\theta +  h^t_{~t} + h^{\varphi}_{~\varphi} + h^{\rho}_{~\rho} 
      -  h^{\theta_i}_{~\theta_i} + h^i_{~i} =0,
\label{relation2-b} 
\end{eqnarray}
Considering the Bianchi identity of Eq.(\ref{bianchi-identity}) leads to 
${\cal H}_\theta=0$.
If we use Eqs.(\ref{relation1-b}) and (\ref{relation2-b}) to decouple 
the last terms in Eqs.(\ref{eq-h-ptr-b}) and (\ref{eq-phi-ptr-b}), 
these lead to
\begin{eqnarray}
&&\bar \nabla^2 (8\delta \phi-h) 
- {4 \over R^2 } h^{i}_{~i}=0,
\label{eq-h-ptr1-b} \\
&&\bar \nabla^2 \delta \phi 
- {2  \over R^2 } h^i_{~i}=0.
\label{eq-phi-ptr1-b} 
\end{eqnarray}
If $h^i_{~i}=0$, then one finds
\begin{equation}
h=8 \delta \phi,~~ \bar \nabla^2 \delta \phi =0.
\label{eq-dilaton3-b}
\end{equation}
However this corresponds trivially to the linearized equation 
for a minimally coupled 
scalar.  If $h^i_{~i} = a \delta \phi$, then 
Eqs.(\ref{eq-h-ptr1-b}) and (\ref{eq-phi-ptr1-b}) lead to
\begin{equation}
\bar \nabla^2 \delta \phi
- {2a  \over R^2 } \delta \phi=0~~~~{\rm with}~~h=6 \delta \phi.
\label{eq-dilaton1-b}
\end{equation}
In order to find $a$, we recall the relation
\begin{equation}
e^{2 \bar \Phi} = g^2 {f_1 \over f_5} = g^2 e^{4 \bar \chi},
\label{relation3-b}
\end{equation}
where $\bar \chi$ is defined through 
$ds_{10}^2 = e^{2 \bar \chi} dx_i^2 + \cdots $.  
The linearized version of Eq.(\ref{relation3-b}) implies
\begin{equation}
\delta \phi = 2 \delta \chi.
\label{relation4-b}
\end{equation}
And then we have 
$g_{ij}=\bar g_{ij} + h_{ij}$ with 
$h^i_{~i} = 8 \delta \chi = 4 \delta \phi$.  Hence 
the final equation for the dilaton in the string frame takes the form
\begin{equation}
\bar \nabla^2 \delta \phi - {8 \over R^2} \delta \phi =0.
\label{eq-dilaton2-b}
\end{equation}
Also this corresponds to the equation for a minimally 
coupled scalar with $l'=2$ on AdS$_3 \times $S$^3$\cite{Lee98}.

\subsection{Greybody Factor}
In order to calculate the greybody factor, 
we have to solve the differential equation(\ref{eq-dilaton2-b}) 
with an appropriate boundary 
condition.  In the conventional study of the BTZ black hole, 
one imposes the boundary condition at infinity to obtain the sensible 
result.  

However we consider the geometry of 
AdS$_3 \times $S$^3 \times $T$^4$ as a whole 
without the boundary condition at spatial infinity.  
First $\delta\phi$ can be decomposed into
\begin{equation}
\delta \phi = e^{-i \omega t} e^{i m \varphi} e^{i K_i x^i} 
  Y_{l'}(\theta_1, \theta_2, \theta_3) \psi(\rho).
\label{psi-wave}
\end{equation}
In the s-wave approximation ($K_i=l'=0$), Eq. (\ref{eq-dilaton2-b}) leads to 
\begin{equation}
\nabla^2_{BTZ} \psi(\rho) - {8 \over R^2} \psi(\rho) =0.
\label{rho-eq}
\end{equation}
Solving (\ref{rho-eq}) with the non-normalizable mode as the boundary 
condition, the absorption cross section with $m=0$ 
is given by\cite{Lee98PRD084022} 
\begin{eqnarray}
\left ( \sigma_{AdS_3 \times S^3}^{\delta \phi}\right )
_{m=0,l'=0} &=&  { 2 \pi R \over 3 } 
\left ( {\pi^3 R^8 \over 64} \right ) \omega 
[\omega^2 + 16 \pi^2 T_L^2][\omega^2 + 16 \pi^2 T_R^2]
 \nonumber  \\
&&~~~~~~\times
{{e^{\omega/T_H} -1} \over
{[e^{\omega/2T_R} -1][e^{\omega/2T_L} -1]}}
\label{abs-l2-b}
\end{eqnarray}
which is the same form as that of a minimally coupled scalar 
with $m=0, l'=2$\cite{Lee98}.
In the low energy limit of $\omega \to 0$, this leads to
\begin{equation}
\left ( \sigma_{AdS_3 \times S^3}^{\delta \phi}\right )
_{m=0,l'=0}^{\omega \to 0} =  { 1 \over 3 } 
\left [ {{\tilde {\cal A}_H^{6D}} \over 4} 
\left ( {r_0 \over R} \right )^4
\right ], ~~ \tilde {\cal A}_H^{6D} = {\cal A}_H^{6D}.
\label{abs-l2-w0-b}
\end{equation}

\section{Discussion}
\label{discussion}
We investigate the dynamical behavior of
the 5D black holes with the dilaton.
Apart from counting the microstates
of black holes, the dynamical behavior is also an important issue
\cite{Hor96,Lee9708,Cal97,Dha96}. This is
so because the greybody factor for
the black hole arises as a consequence
of scattering of a test field off the gravitational
potential barrier surrounding the horizon.
That is, this is an effect of spacetime curvature. Together with the
Bekenstein-Hawking entropy\cite{Sfe98NPB179,Car95,Str97}, 
this seems to be a strong hint of a deep
and mysterious connection between curvature and statistical mechanics.

First we study the greybody factor of the dilaton for a 5D black hole 
in M$_5 \times $S$^1 \times $T$^4$ with asymptotically flat space.
It was understood that the greybody factor calculation makes 
sense when one defines asymptotic region as in Sec.\ref{ads}
\cite{Gub98PLB105,Wit98ATMP253}.
Hence it may not be possible in global AdS$_3$ because there is no 
asymptotic state corresponding to particle at infinity in AdS$_3$.
However, we calculate the absorption cross section of the 
dilaton in AdS$_3 \times $S$^3 \times $T$^4$.  
Actually we use the matching of the AdS region to 
asymptotically AdS$_3$ to obtain the greybody factor.
In this case we don't require any boundary condition\cite{Lee98PLB83}, because 
the non-normalizable mode is used to calculate the greybody factor. 
This mode acts as a fixed background and its boundary value 
acts as a source for operator with (2,2) in the boundary theory. 

Two of authors(Lee and Myung) showed that the s-wave greybody
factor of a minimally coupled scalar for 
the AdS$_3 \times $S$^3 \times $T$^4$ has the
same form as the one for M$_5 \times $S$^1 \times $T$^4$ 
in the dilute gas approximation\cite{Lee98}. 
In this case, a free field was used for a calculation\cite{Hor96,Dha96}. 
In the present calculation,
we use the 10D dilaton(a fixed scalar) as a test field\cite{Lee9708,Cal97}.  
We find the same form of absorption cross section
for both two cases as
\begin{equation}
\left ( \sigma_{M_5 \times S^1}^{\delta \phi} \right )_{l=0,K_5=0} =
3 \left ( \sigma_{AdS_3 \times S^3}^{\delta \phi} \right )_{m=0,l'=0},
\label{absorption}
\end{equation} 
upto the constant 3.
This means that the AdS$_3$-calculation recovers a result of 
5D black hole. 

In conclusion, we study 
the cross section for the dilaton within two supergravity solutions. 
The first is on the M$_5 \times $S$^1 \times $T$^4$ ( 5D black hole 
with asymptotically flat space) and the second is on  
AdS$_3 \times $S$^3 \times $T$^4$( the BTZ black hole
with asymptotically AdS). 
Hence the near horizon geometries of two solutions are the 
same as AdS$_3$, but the asymptotic structures are quite different.
The latter calculation is based on the non-normalizable mode 
which is very useful for both the matching procedure and bounadry condition 
at infinity of the AdS$_3$\cite{Bal99PRD046003}.

\section*{Acknowledgement}
This work was supported in part by the Basic Science Research Institute 
Program, Minstry of Education, Project NOs. BSRI-98-2413 and 
BSRI-98-2441. 

\appendix
\section{The Calculation of $\bar R_{MN}$ in the Einstein frame}
\label{RMN}
The explicit form of $\bar R_{MN}$ with 
$r_1=r_5 \gg r_0,r_n$ is given by
\begin{eqnarray}
\bar R_{tt}&=&
  {2 d ( f_n^2 r_5^4 d + r_n^4 f_5^2 d + f_n^2 r_5^2 r_0^2 f_5 +
        r_n^2 f_5^2 r_0^2 f_n ) \over r^6 f_5^4 f_n^3 },
\\
\bar R_{xx}&=&
  -{2 ( f_n^2 r_5^4 d - r_n^4 f_5^2 d + f_n^2 r_5^2 r_0^2 f_5 -
        r_n^2 f_5^2 r_0^2 f_n ) \over r^6 f_5^4 f_n },
\\
\bar R_{rr}&=&
  -{2 ( f_n^2 r_5^4 d + r_n^4 f_5^2 d + f_n^2 r_5^2 r_0^2 f_5 +
        r_n^2 f_5^2 r_0^2 f_n ) \over r^6 f_5^2 f_n^2 d },
\\
\bar R_{\theta_1 \theta_1}&=&
  {2 (  r_5^4 d + r_0^2 f_5 r_5^2 - r_0^2 f_5^2 r^2 +
        r^4 f_5^2 - r^4 f_5^2 d ) \over r^4 f_5^2 },
\\
\bar R_{\theta_2 \theta_2}&=&
  {2 \sin^2\theta_1(  r_5^4 d + r_0^2 f_5 r_5^2 - r_0^2 f_5^2 r^2 +
        r^4 f_5^2 - r^4 f_5^2 d ) \over r^4 f_5^2 },
\\
\bar R_{\theta_3 \theta_3}&=&
  {2 \sin^2\theta_1 \sin^2 \theta_2 
       (  r_5^4 d + r_0^2 f_5 r_5^2 - r_0^2 f_5^2 r^2 +
        r^4 f_5^2 - r^4 f_5^2 d ) \over r^4 f_5^2 },
\end{eqnarray}
Here one finds
\begin{equation}
\bar R = -{ 2(r_0^2 + r_n^2) r_n^2 \over 
r^6+r^4r_5^2 + 2 r^4 r_n^2 +2 r^2 r_n^2r_5^2 +  r^2 r_n^4 + r_n^4 r_5^2 }
\simeq 0
\end{equation}
in the near horizon.

\section{The block diagonal form of graviton for M$_5 \times $S$^1 \times $T$^4$
(AdS$_3 \times $S$^3 \times $T$^4$)}
\label{hMN}
\begin{equation}
h_{MN} = \left [
\begin{array}{ccccccc}
h_{tt} & h_{t x_5} & h_{tr} & &&&\\
h_{x_5 t} & h_{x_5 x_5} & h_{x_5 r} & &0 &&0 \\
h_{rt} & h_{r x_5} & h_{rr} & &&&\\
&&&h_{\theta_1 \theta_1} & h_{\theta_1 \theta_2 } & h_{\theta_1 \theta_3} &\\
&0 &&h_{\theta_2 \theta_1} & h_{\theta_2 \theta_2 } & h_{\theta_2 \theta_3} &0\\
&&&h_{\theta_3 \theta_1} & h_{\theta_3 \theta_2 } & h_{\theta_3 \theta_3} &\\
&0&&&0 & &h_{ij}
\end{array}
\right ]
\end{equation}


\begin{references}

\bibitem{Mal98ATMP231} J. Maldacena, Adv. Theor. Math. Phys. {\bf 2}, 
  231(1998), hep-th/9711200. 
\bibitem{Gub98PLB105}
  S.S. Gubser, I.R. Klebanov and A.M. Polyakov, 
  Phys. Lett. {\bf B428}, 105(1998), hep-th/9802109.
\bibitem{Wit98ATMP253}
  E. Witten, Adv. Theor. Math. Phys. {\bf 2}, 253(1998), hep-th/9802150.
\bibitem{Hor98PRL4116}
  G.T. Horowitz and H. Oguri, Phys. Rev. Lett. {\bf 80}, 4116(1998), hep-th/9802116;
  E. Witten, Adv. Theor. Math. Phys. {\bf 2}, 505(1998), hep-th/9803131;
  G.T. Horowitz and S. Ross, JHEP 9804, 015(1998), hep-th/9803085.
\bibitem{Hyu97} S. Hyun, hep-th/9704005.
\bibitem{Sfe98NPB179} 
  K. Sfetsos and K. Skenderis, Nucl. Phys. {\bf B517}, 179(1998), hep-th/9711138.
\bibitem{Ban92} 
M. Banados, C. Teitelboim and A. Zanelli, Phys. Rev. Lett.
{\bf 69}, 1849(1992);
S. Carlip, Class. Quant. Grav. {\bf 12}, 2853(1995).
\bibitem{Hor93} 
G. Horowitz and D. Welch, Phys. Rev. Lett. {\bf 71}, 328(1993);
N. Kaloper, Phys. Rev. {\bf D48}, 2598(1993);
A. Ali and A. Kumar, Mod. Phys. Lett. {\bf A8}, 2045(1993).
\bibitem{Car95} S. Carlip, Phys. Rev. {\bf D51}, 632(1995);
{\bf D55}, 878(1997).
\bibitem{Str97}
  A. Strominger, JHEP 9802, 009(1998), hep-th/9712251;
  D. Birmingham, I. Sachs ans S. Sen, Phys. Lett. {\bf B424}, 275(1998), hep-th/9801019;
  N. Kaloper, Phys. Lett. {\bf B434}, 285(1998), hep-th/9804062.
\bibitem{Beh98PLB310}
  K. Behrndt, I. Brunner, and I. Gaida, 
  Phys. Lett. {\bf B432}, 310(1998), hep-th/9804159.
\bibitem{Lee98}
  H.W. Lee and Y.S. Myung, Phys. Rev. {\bf D58}, 104013(1998), hep-th/9804095.
\bibitem{Hor96}
  G.T. Horowitz, J.M. Maldacena and A. Strominger, 
     Phys. Lett. {\bf B383}, 151(1996);
  F. Dowker, D. Kastor and J. Traschen, 
    Phys. Rev. {\bf D58}, 124025(1998), hep-th/9702109.
\bibitem{Gre93}
  R. Gregory and R. Laflamme, Phys. Rev. Lett. {\bf 70}, 2837(1993).
\bibitem{Lee9801} 
 H.W. Lee, N.J. Kim, Y.S. Myung, and J.Y. Kim, Phys. Rev. {\bf D57}, 7361(1998).
\bibitem{Lee98PRD084022} 
 H.W. Lee, N.J. Kim, and Y.S. Myung, Phys. Rev. {\bf D58}, 084022(1998), 
              hep-th/9803080.
\bibitem{Lee9708}
  H.W. Lee, Y.S. Myung and J.Y. Kim, Phys. Rev. {\bf D58}, 104006(1998), 
     hep-th/9708099.
\bibitem{Cal97}
  C. Callan, S. Gubser, I.G. Klebanov and A. Tseytlin, Nucl. Phys.
    {\bf B489}, 65(1997);
  M. Krasnitz and I.G. Klebanov, 
    Phys. Rev. {\bf D56}, 2173(1997), hep-th/9703216.
\bibitem{Dha96}
  S.D. Mathur, hep-th/9704050;
  S.S. Gubser, Phys. Rev. {\bf D56}, 4984(1997), hep-th/9704195;
  A. Dhar, G. Mandal and S. Wadia, Phys. Lett. {\bf B388}, 51(1996);
  S. Das, G. Gibbons and S. Mathur, Phys. Rev. Lett. {\bf 78}, 417(1997).
\bibitem{Lee98PLB83}
  H.W. Lee, N.J. Kim, and Y.S. Myung, Phys. Lett. {\bf B441}, 83(1998); 
   hep-th/9803227.
\bibitem{Bal99PRD046003}
  V. Balasubramanian, P. Kraus, and A. Lawrence, 
       Phys. Rev. {\bf D59}, 046003(1999); hep-th/9805171.
\end{references}
\end{document}